\documentclass[%
 reprint,
 superscriptaddress,
 preprintnumbers,
 nofootinbib,
 amsmath,amssymb,
 aps,
 prl,
]{revtex4-1}
\usepackage{psfrag,slashed,cancel,array,graphicx,todonotes,hyperref}
\usepackage{subfigure}
\usepackage[labelfont=bf,labelsep=quad,justification=justified]{caption}
\usepackage[utf8]{inputenc}
\usepackage{mathtools}
\usepackage{mathrsfs}
\usepackage{multirow}
\usepackage{braket}
\usepackage{slashed}
\usepackage{booktabs} 
\usepackage[normalem]{ulem}
\usepackage{slashed}
\hypersetup{pdftitle={Charge-Charge Correlation in the Back-to-Back limit},pdfcreator={},linkcolor=[rgb]{0.15,0.35,0.75},colorlinks=true,citecolor=[rgb]{0.675,0,0.2},urlcolor=[rgb]{0.15,0.35,0.65}}

\newcommand{\refcite}[1]{ref.~\cite{#1}}
\newcommand{\refscite}[1]{refs.~\cite{#1}}

\newcommand{\eq}[1]{eq.~\eqref{eq:#1}}
\newcommand{\eqs}[2]{eqs.~\eqref{eq:#1} and \eqref{eq:#2}}

\newcommand{\fig}[1]{figure~\ref{fig:#1}}

\newcommand\myeq{\,\,\,\stackrel{\mathclap{{\footnotesize\mbox{$z\to 1$}}}}{=}\,\,\,}

\newcommand{\df}{\mathrm{d}}
\newcommand{\img}{\mathrm{i}}

\newcommand{\zb}{\bar z}

\newcommand{\bn}{{\bar n}}
\newcommand{\bq}{{\bar q}}

\newcommand{\cJ}{\mathcal{J}}
\newcommand{\cK}{\mathcal{K}}
\newcommand{\cL}{\mathcal{L}}

\newcommand{\cN}{\mathcal{N}}
\newcommand{\cO}{\mathcal{O}}

\newcommand{\qt}{{\vec q}_T}
\newcommand{\bt}{{\vec b}_T}

\newcommand{\JQQC}{J^{\rm QQC}}
\newcommand{\JEEC}{J^{\rm EEC}}

\newcommand{\as}{\alpha_s}
\newcommand{\GammaC}{\Gamma_{\rm cusp}}
\newcommand{\nn}{\nonumber}

\newcommand{\lqcd}{\Lambda_\mathrm{QCD}}

\newcommand{\born}{ \hat \sigma_0}

\def\beq{\begin{equation}}
\def\eeq{\end{equation}}
\def\bea{\begin{eqnarray}}
\def\eea{\end{eqnarray}}

\AtBeginDocument{
\heavyrulewidth=.08em
\lightrulewidth=.05em
\cmidrulewidth=.03em
\belowrulesep=.65ex
\belowbottomsep=0pt
\aboverulesep=.4ex
\abovetopsep=0pt
\cmidrulesep=\doublerulesep
\cmidrulekern=.5em
\defaultaddspace=.5em
}

\def\cO{\mathcal{O}}

\def\cN{\mathcal{N}}

\def\be{\begin{equation}}
\def\ee{\end{equation}}

\begin{document}

\preprint{CERN-TH-2025-150}

\title{On the Edge of Safety: Charge-Charge Correlation in the Back-to-Back Limit}

\author{Pier Francesco Monni}
\email{pier.monni@cern.ch}
\affiliation{CERN, Theoretical Physics Department, CH-1211 Geneva 23, Switzerland}
\author{Gherardo Vita}
\email{gherardo.vita@cern.ch}
\affiliation{CERN, Theoretical Physics Department, CH-1211 Geneva 23, Switzerland}
\author{Zhen Xu}
\email{zhen.xu@zju.edu.cn}
\affiliation{Zhejiang Institute of Modern Physics, Department of Physics, Zhejiang University, Hangzhou
310027, China}
\author{Hua Xing Zhu}
\email{zhuhx@pku.edu.cn}
\affiliation{School of Physics, Peking University, Beijing 100871, China}
\affiliation{Center for High Energy Physics, Peking University, Beijing 100871, China}

\begin{abstract}
We investigate the Charge-Charge Correlation (QQC) in electron-positron annihilation as a probe of charge dynamics in Quantum Chromodynamics. While generally divergent beyond leading order, we show that the QQC is infrared and collinear safe in the back-to-back limit, a property that we dub leading-power safety. This enables an analytic perturbative treatment that bypasses the reliance on non-perturbative track or fragmentation functions. Using Soft-Collinear Effective Field Theory, we derive a factorization theorem and determine its logarithmic behavior analytically up to four loops in QCD, uncovering remarkable connections with the Energy-Energy Correlation (EEC). 
Prior to this work, the behavior of the QQC beyond leading order was entirely unknown; we now establish its resummation to next-to-next-to-next-to-next-to-leading logarithmic (N$^4$LL) accuracy, placing it among the observables with the highest perturbative precision alongside the EEC. 
Our results are validated through a numerical analysis using \texttt{Event2}, exhibiting excellent agreement with the predicted singular terms.
This work establishes the QQC as a novel, calculable probe of charge dynamics and unveils a new window into the inner workings of non-Abelian gauge theories.
\end{abstract}

\maketitle

\paragraph*{Introduction.---}

Understanding the internal structure of hadronic events in Quantum Chromodynamics~(QCD) has long been a central objective in high-energy physics.  
Among the most powerful probes of this structure are event-shape observables and correlation functions, which can reveal the detailed flow of quantum numbers such as energy and charge.  

In this work, we investigate the Charge-Charge Correlation (QQC) observable, defined in electron-positron annihilation as the pairwise angular correlation between the charges of all final-state particles.  
The QQC provides a direct measure of how electric charge is distributed across an event and how it correlates between widely separated regions of phase space.  
Specifically, the QQC can be defined as
\be
\text{QQC}(\chi) = \sum_{a,b} \int \df \sigma_{e^+e^- \to a,b + X} \frac{Q_a Q_b}{\sigma_{\text{tot}}} \delta(\cos \chi - \vec{n}_a \cdot \vec{n}_b) \, ,
\ee
where $Q_i$ is the electric charge of particle $i$, the sum runs over all final-state particles, $\hat{n}_a$ represents the light-like direction of the parton $a$, and $\sigma_{\text{tot}}$ is the total cross section.
\begin{figure}[h]
  \centering
  \includegraphics[width=0.35\textwidth]{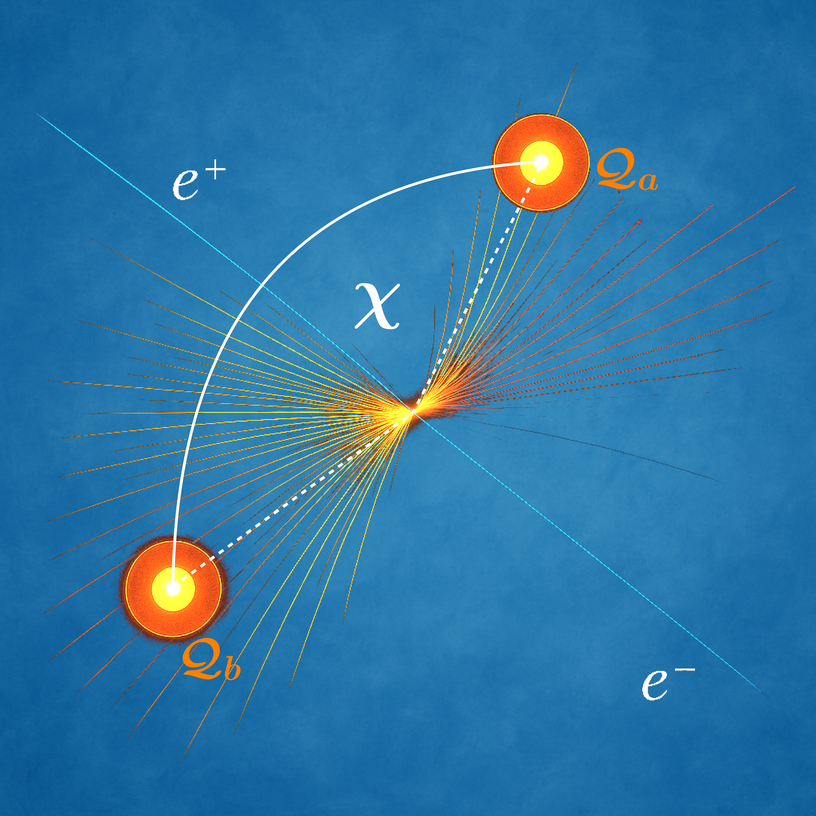} 
  \caption{Graphical illustration of the charge-charge correlation in electron-positron annihilation.}
  \label{fig:QQCimage}
\end{figure}%
Formally, the charge can be defined as the light transform of the electromagnetic current \cite{Ore:1979ry,Sveshnikov:1995vi,Korchemsky:1997sy,Korchemsky:1999kt,Belitsky:2001ij, Lee:2006nr,Hofman:2008ar,Chicherin:2020azt,Cuomo:2025pjp}:
\begin{align} \label{Q-flow}
\mathcal Q(\vec n) = \lim_{r\to\infty} r^2\, \int_{-\infty}^\infty dt \,  n^i\, J_{i }(t,r\vec n)\,.
\end{align}
Therefore, the QQC too can be defined in terms of  a double insertion of such charge flow operators as  
\begin{equation}
\label{eq:QQCformal}
\mathrm{QQC}(\chi) = \frac{1}{\sigma_{\text{tot}}} \int\! d^4 x\, e^{iqx}  \langle  J(x) \, \mathcal Q(\vec n_a)  \mathcal Q(\vec n_b)\, J(0)\rangle_W \,, 
\end{equation}
with $\vec n_a \cdot \vec n_b = \cos\chi $, and $q$ the total momentum of the hadronic state.
Consequently, the QQC is related to the four-point Wightman correlation function of the electromagnetic current operator, providing a connection to fundamental operator structures within the theory and highlights its potential sensitivity to the intricate charge dynamics of the underlying gauge theory.
This connection places the QQC within a broader program aimed at understanding collider observables through the language of correlation functions, a perspective that has recently gained prominence across a range of theoretical studies \cite{Hofman:2008ar,Belitsky:2013bja,Belitsky:2013ofa,Belitsky:2013xxa,Komargodski:2016gci,Henn:2019gkr,Kologlu:2019mfz,Korchemsky:2019nzm,Chicherin:2020azt,Chen:2020vvp,Chang:2020qpj,Chen:2020adz,Korchemsky:2021okt,Chen:2021gdk,Chen:2022jhb,Chang:2022ryc,Caron-Huot:2022eqs,Lee:2022uwt,Lee:2023npz,Chen:2023wah,Chen:2023zzh,Chen:2024nyc,He:2024hbb,Cuomo:2025pjp}, see \cite{Moult:2025nhu} for a recent review.
Despite its conceptual appeal, the QQC has received limited theoretical attention because it is known to be infrared and collinear (IRC) unsafe beyond leading order (LO) in perturbative QCD~\cite{Lee:1981mf}. This has historically precluded precision predictions for the QQC, necessitating reliance on non-perturbative models of hadronization via track or fragmentation functions, see e.g. \cite{Chang:2013rca,Li:2019dre,Kang:2020fka}. 
\begin{figure}
  \centering
  \includegraphics[width=0.35\textwidth]{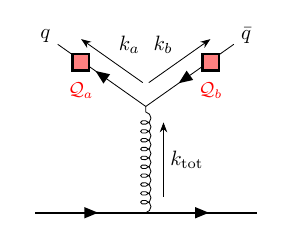} 
  \caption{Charge correlation of a correlated $q\bar{q}$ soft pair.}
  \label{fig:gtoqqbar}
\end{figure}

In this Letter, we demonstrate that the QQC becomes IRC safe, and thus perturbatively calculable, in the back-to-back kinematic limit, where $\chi \to \pi$, corresponding to the $z \equiv (1-\cos\chi)/2 \to 1$ regime. We will refer to this property as \emph{leading power (LP) safety}. This observation is used to derive a state of the art prediction for this observable at lepton colliders, enabling high-precision phenomenological studies of charge dynamics in QCD.

\paragraph*{Physics of QQC in the back-to-back limit.---}
The back-to-back limit of the QQC is characterized by the propagation of two streams of energetic particles with the charge detectors placed on pairs of charged particles belonging to opposite streams. Correlations of particles not aligned with the energetic streams are subdominant, and hence suppressed by powers of $(1-z)$ for $z\to 1$. Beyond LO in QCD, there are two sources of IRC unsafety that plague the QQC, both stemming from placing a detector on quarks that originate from the branching of a very soft gluon, as shown in \fig{gtoqqbar}. This process introduces soft charges into the final state in a manner that disrupts the cancellations of real and virtual divergences necessary for IRC safety.
The first source arises when the resulting $q\bar{q}$ pair propagates along one of the two streams, and are correlated with particles belonging to the opposite stream. This configuration is protected by charge conservation, that ensures that the resulting contribution to the QQC vanishes upon summing over the $q$ and the $\bar{q}$. 
The second source of IRC unsafety is due to a soft gluon with a wide angle with respect to the large energy flow that splits into a $q\bar{q}$ pair that fly into opposite directions. While this configuration is not protected by charge conservation, we observe that the corresponding squared amplitude is exponentially suppressed in the back-to-back limit, that is when the rapidity separation $\Delta\eta_{q\bar{q}}$ between the $q\bar{q}$ pair is large. The squared amplitude for the splitting in \fig{gtoqqbar} scales as~\cite{Catani:1999ss}
\begin{equation}
    |{\cal M}_{q\bar{q}}|^2 \sim (8\pi\mu^{2\epsilon}\alpha_s)^2 C_F \frac{|\vec{k}_{T,q}- \vec{k}_{T,\bar{q}}|^2}{|\vec{k}_{T,q}|^3|\vec{k}_{T,\bar{q}}|^3} e^{-3|\Delta\eta_{q\bar{q}}|}\,,
\end{equation}
with $\vec{k}_{T,i}$ denoting the transverse momentum w.r.t. the direction of the energetic particles. This observation crucially implies that the divergence is suppressed in the back-to-back limit by powers of $(1-z)$, hence allowing for the application of perturbative methods to the calculation of QQC in this regime. Using the framework of Soft-Collinear Effective Theory (SCET)~\cite{Bauer:2000ew, Bauer:2000yr, Bauer:2001ct, Bauer:2001yt} we now establish for the first time a factorization theorem in QCD for the leading power asymptotic of QQC in this limit.
%

\paragraph*{Factorization theorem in SCET.---}
Our theoretical framework starts from the description of di-hadron production in the limit of small transverse momentum $q_T$ of the virtual photon in the back-to-back frame of the two hadrons. The factorization for this regime was established by Collins and Soper~\cite{Collins:1981uk, Collins:1981va,Collins:2011zzd} and reads
\begin{align} \label{eq:qT_fact_q}
 \frac{\df\sigma_{e^+ e^- \to h_1 h_2 + X}}{\df x_1 \df x_2 \df^2\qt} &
 =  \int\!\frac{\df^2\bt}{(2\pi)^2} \, e^{\img \qt \cdot \bt} H_{q\bar q}(Q) D_{h_1 / q}(x_1, b_T) \nn\\ &\qquad \times   D_{h_2 / \bar q} (x_2, {b_T}) S_q(b_T)
\,,\end{align}
up to power corrections in $\cO(q_T^2/Q^2)$.
In the above expression, a summation over all quark flavors $q$ is understood. The hard function $H_{q\bar q}$ captures the virtual corrections to the tree-level process $e^+ e^- \to q \bar q$, and it is known to N$^4$LO in QCD~\cite{Lee:2022nhh,Chakraborty:2022yan}. The function $ D_{h/q}$ represents the transverse-momentum-dependent fragmentation function (TMDFF), describing the probability for a quark $q$ to hadronize into a hadron $h$. The soft function $S_q$ accounts for the net transverse momentum generated by soft‐gluon radiation in the event. It is known to N$^3$LO~\cite{Li:2016ctv} and it coincides with the soft function employed in $q_T$ resummation for $pp$ collisions~\cite{Collins:2004nx}, as it is independent of the orientation of the associated Wilson lines~\cite{Moult:2018jzp,Zhu:2020ftr}.

We can then use eq.~\eqref{eq:qT_fact_q} to write the QQC in the back-to-back limit as
\begin{align}\label{eq:factQQCreltodihadron}
\frac{\df\sigma^{e^+e^-}_\mathrm{QQC}}{\df z} &\myeq  \int_0^1 \df x_a \int_0^1 \df x_b \int \df^2 \qt  \delta\left(1 - z - \frac{\qt^{\,2}}{Q^2} \right) \nn\\&\qquad \sum_{h_a,h_b} Q_{h_a}Q_{h_b} \frac{\df\sigma_{e^+ e^- \to h_a h_b + X}}{\df x_a \df x_b \df^2\qt}\,,   
\end{align}
up to power corrections of $\cO((1-z)^0)$. In the form of \eq{factQQCreltodihadron}, the QQC still involves the non-perturbative TMDFF $D_{h_a / q}$. In the following, we prove that \eq{factQQCreltodihadron} is actually independent of them, in line with the expectation for an IRC safe observable. We start by noting that for perturbative values of $q_T$, we can derive an operator product expansion (OPE) for the TMDFF \cite{Aybat:2011zv,Collins:2011zzd}. At leading power in $\cO(b_T^2 \lqcd^2)$ we can match the TMDFF onto the longitudinal FF $d_{h/q}$ as
\begin{align} \label{eq:TMDFF_matching}
 D_{h / q}(x, b_T)
 = \sum_i \int_{x}^1\! \frac{\df t}{t} 
   \cK_{q i}(t, b_T)\, d_{h/i}\Bigl(\frac{x}{t}\Bigr),\!\!
\end{align}
where $\cK_{q i}$ is a perturbative matching kernel and the sum runs over all parton flavors $i$ including gluons. These matching kernels are well-known objects, and have been calculated up to N$^3$LO in \refscite{Ebert:2020qef,Luo:2020epw}. Taking into account the integral over the energies of the tagged hadron and its charge weight from \eqref{eq:factQQCreltodihadron} we can factorize the dependence on the longitudinal FF obtaining 
{\small \begin{align} \label{eq:KFfactorization}
\JQQC_q &\equiv \sum_{h_a} \int_0^1\! \df x_a Q_{h_a} D_{h_a / q}(x_a, b_T)\\
 &=  \sum_{h_a,i} \int_0^1\! \df x_a \int_{x_a}^1\! Q_{h_a}\frac{\df z_a}{z_a} 
    \cK_{q i}(z_a, b_T) d_{h_a/i}\Bigl(\!\frac{x_a}{z_a}\!\Bigr)\nn\\
 &=  \sum_{i}  \int_{0}^1\! \df z_a 
    \cK_{q i}(z_a, b_T) \left[\sum_{h_a}\int_0^1\! \df \tau_a Q_{h_a} d_{h_a/i}(\tau)\right]\nn\\
     &=  \sum_{i}  \int_{0}^1\! \df z_a 
     \cK_{q i}(z_a, b_T) Q_i\,,
\end{align}}
where we used the charge sum rules for quark fragmentation functions \cite{Collins:1981uw}.
Therefore, we have demonstrated that leading power asymptotic for the QQC does not depend on non-perturbative objects. 
We can thus rewrite \eqref{eq:factQQCreltodihadron} in terms of purely perturbative objects obtaining our factorization formula for QQC in the back-to-back limit
\begin{align} \label{eq:QQC_fact_thm}
 &{\rm QQC}(z)
 = \frac{\born}{2} \sum_{q} H_{q\bar q}(Q,\mu)\!\! \int_0^\infty\!\!\! \df (b_T Q)^2 J_0\bigl(b_T Q \sqrt{\zb}\bigr)\nn\\&~~~~\times
   \JQQC_q\Bigl(b_T, \mu, \frac{b_T Q}{\upsilon}\Bigr) \JQQC_\bq\Bigl(b_T, \mu, \upsilon b_T Q \Bigr) [ 1 + \cO(\zb) ]
\,,\end{align}
where $\zb \equiv 1-z$, $H_{q\bq}$ is the same hard function as in \eq{qT_fact_q},
$J_0(x)$ is the $0$-th Bessel function of the first kind, and $\JQQC_q$ are the QQC jet functions~\eqref{eq:KFfactorization}, that appear in this context for the first time. In eq.~\eqref{eq:QQC_fact_thm} $\mu$ denotes the renormalization scale and $\upsilon$ denotes the rapidity regularization scale in the pure regularization scheme~\cite{Ebert:2018gsn,Duhr:2022yyp} in which the soft function is equal to one.

\paragraph*{The QQC Jet Function.---}
 The existence of the QQC jet function~\eqref{eq:KFfactorization} is far from trivial. Indeed, each of the renormalized matching kernels $\cK_{ij}(x,\bt)$ entering in its definition diverge when integrated over the whole range of energies $x \in [0,1]$. However, the integral defining $\JQQC$ is well defined. To show this we make use of the sign difference in the charge when summing over quarks and antiquarks to obtain
\begin{align}\label{eq:JQQCdef}
        \JQQC_q(b_T) 
        &= \sum_{q^\prime} Q_{q^\prime} \int_{0}^1\df x \left[\cK_{q q^\prime}(x,b_T) -\cK_{q\bar{q}^\prime}(x,b_T)\right]    \,,
\end{align}
where $q^\prime$ runs over all flavors (including $q$), but excludes the gluon and the anti-quarks (or the quarks in case $q$ is an anti-quark). Non-trivially, the $x\to0$ limit is regulated by the universality of the Regge limit of quark-antiquark matching kernels, i.e. the fact that
\beq
\lim_{x\to 0}\left[\cK_{qq^\prime}(x,b_T)-\cK_{q\bar{q}^\prime}(x,b_T)\right] = \cO(x^0)\,,
\eeq
which has been explictly shown to 3 loops in \refscite{Ebert:2020qef,Luo:2020epw}.
On the other hand, for $x\to1$ eq.~\eqref{eq:JQQCdef} is finite due to the standard UV regularization in SCET of the matching kernels.
Using \eq{JQQCdef} and the result for the TMDFF calculated in \cite{Ebert:2020qef,Luo:2020epw}, we obtain the QQC jet function to N$^3$LO and we make them available in ref.~\cite{vita_2025_16658462}.
The anomalous dimension of QQC jet function is completely dictated by the anomalous dimension of the hard function by the consistency of the factorization theorem. Moreover, since $\JQQC$ represents the only difference in the factorization theorem compared to the EEC SCET factorization in the back-to-back limit \cite{Moult:2018jzp,Ebert:2020sfi,Duhr:2022yyp} (see also \cite{Collins:1981uk,deFlorian:2004mp,Aglietti:2024xwv}) the consistency of both factorizations implies
\be\label{eq:muRGEConsistency}
\mu \frac{\df}{\df\mu}\ln\left[\JQQC_q\Bigl(b_T,\mu,\frac{b_T Q}{\upsilon}\Bigr)/\JEEC_q\Bigl(b_T,\mu,\frac{b_T Q}{\upsilon}\Bigr)\right] = 0\,,
\ee
as well as
\be\label{eq:nuRGEConsistency}
\upsilon \frac{\df}{\df \upsilon}\ln\left[\JQQC_q\Bigl(b_T,\mu,\frac{b_T Q}{\upsilon}\Bigr)/\JEEC_q\Bigl(b_T,\mu,\frac{b_T Q}{\upsilon}\Bigr)\right] = 0\,,
\ee
where
\begin{align}\label{eq:JEECCdef}
        \JEEC_q\Bigl(b_T,\mu,\frac{b_T Q}{\upsilon}\Bigr) 
        &= \sum_{i} \int_{0}^1\df x \, x \, \cK_{q i}(x,b_T,\mu,\frac{b_T Q}{\upsilon})    \,,
\end{align}
is the EEC jet function \cite{Moult:2018jzp,Ebert:2020sfi,Duhr:2022yyp}.
Using the newly determined expression for $\JQQC$ and the three loop results for $\JEEC$ from \refcite{Ebert:2020sfi} we verified explicitly that both \eqs{muRGEConsistency}{nuRGEConsistency} hold to three loops, providing a strong analytic cross-check on the three loop determination of $\JQQC$ in this Letter. 
%
The corresponding RGEs for the ingredients of the factorization theorem are given in refs.~\cite{Ebert:2020sfi,Duhr:2022yyp} and we report them in ref.~\cite{supplemental} for completeness. Their solution is expressed in terms of the anomalous dimensions and boundary conditions leading to the resummed formula for the QQC in the back-to-back limit
{ \begin{align} \label{eq:QQC_resummed}
 &\frac{\df\sigma}{\df z} 
 = \frac{\born}{2} \int_0^\infty \!\!\df (b_T Q)^2 \, J_0\bigl(b_T Q \sqrt{1-z}\bigr)H_{q\bq}(Q,\mu_H)
 \\ &
 \times\cJ^{\rm QQC}_q\Bigl(b_T, \mu_J, \frac{Q b_T}{ \upsilon_n}\Bigr) \cJ^{\rm QQC}_\bq\Bigl(b_T, \mu_J, Q b_T \upsilon_\bn \Bigr)  \Bigl(\frac{\upsilon_n}{\upsilon_\bn}\Bigr)^{ \frac12\gamma_r^q(b_T, \mu_J)}
 \nn\\ &\times
 \exp\left[ 4\int_{\mu_J}^{\mu_H} \frac{\df\mu'}{\mu'} \GammaC^q[\as(\mu^\prime)] \ln\frac{\mu^\prime}{Q} -  \gamma^q_H[\as(\mu^\prime)]\right]
\nn\,.\end{align}}
Here $\mu_J$, $\mu_H$, $\upsilon_n$, and $\upsilon_{\bar n}$ are the scales at which the boundary conditions of the RGEs are evaluated, to be taken of the order of the canonical scales.
The determination of the QQC jet function to N$^{3}$LO allows for the evaluation of \eq{QQC_resummed} to N$^4$LL accuracy.

\paragraph*{Analytic and Numerical Validation---}
At LO, $(\cO(\alpha_s))$, the QQC is IRC safe, we can therefore obtain the back-to-back limit from the expansion of its full angle expression obtaining~\cite{Chicherin:2020azt,Chen:2025xxx}
\begin{align}
\frac{\mathrm{QQC}^{\rm LO}(z)}{\frac{\alpha_s}{4\pi}} \myeq\,\,& C_{F} \left( 8\cL_1(\zb) +12\cL_0(\zb) + (8 + 8\zeta_2)\delta(\zb) \right)\,
\end{align}
which matches the $\mathcal{O}(\alpha_s)$ expansion of our formula, including for the contact terms proportional to $\delta(\zb)$.

Furthermore, a powerful consistency check can be performed by considering $\mathcal{N}=4$ supersymmetric Yang-Mills (SYM) theory.
In $\mathcal{N}=4$ SYM, the operator defining the QQC is identical to that of the energy-energy correlator (EEC), as both relate to correlation functions of operators $\langle \cO(x_1) \cO(x_2) \cO(x_3) \cO(x_4) \rangle $ within the stress-tensor supermultiplet~\cite{Hofman:2008ar, Belitsky:2013xxa, Belitsky:2013bja, Korchemsky:2021okt}. 
Consequently, we have
\be
\mathrm{QQC}_{\rm \cN=4}(z) = \mathrm{EEC}_{\rm \cN=4}(z) \,.
\ee
To verify this relation, we apply the principle of maximal transcendentality~\cite{Kotikov:2002ab} to the QQC and substitute $C_F\to C_A,n_f \to C_A$ to obtain its $\cN=4$ form from QCD and compare the known results for the EEC in the back-to-back limit in $\mathcal{N}=4$ SYM \cite{Korchemsky:2019nzm,Ebert:2020sfi,Moult:2022xzt}. We carried out this procedure to 3 loops, finding perfect agreement between the two results. This yields an independent prediction for the leading transcendental terms in the back-to-back limit, providing a valuable cross-check against our results obtained via our  factorization theorem.

\begin{figure}
  \centering
\includegraphics[width=0.49\textwidth,trim= 0.8cm 0 2.5cm 0.8cm, clip]{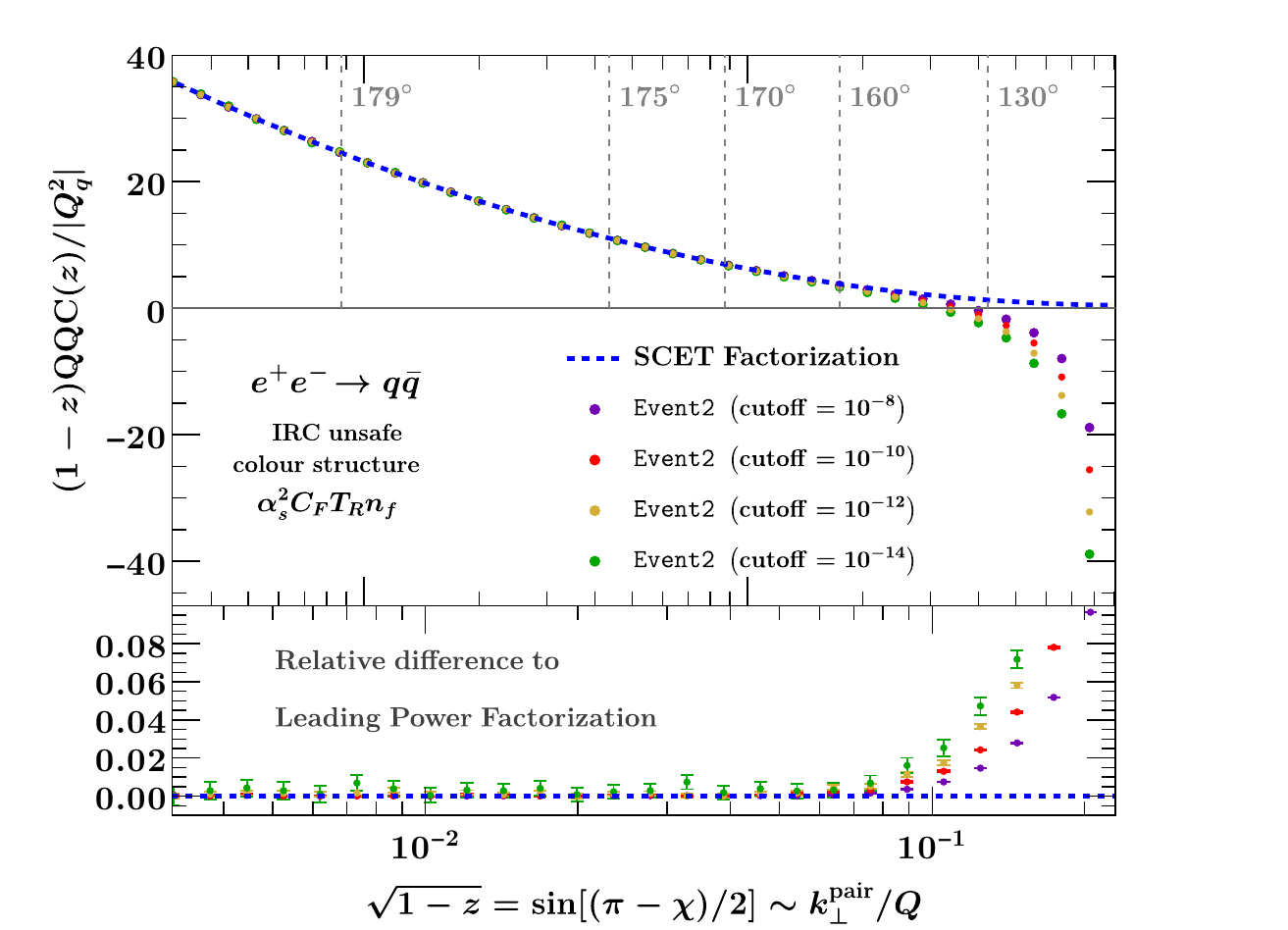} 
  \caption{Numerical validation of the leading power factorization in \eq{QQC_fact_thm} with \texttt{Event2}~\cite{Catani:1996vz}. In the back-to-back limit the numerical result is cutoff independent and matches the analytic prediction of the SCET factorization theorem. Away from the limit, the numerical results deviate from the leading power prediction and show a stronger cutoff dependence due to the IRC unsafety of the QQC at subleading powers in $1-z$.}
  \label{fig:event2_backtoback}
\end{figure}
To ensure our factorization framework accurately describes the contributions responsible for the IRC unsafety of the QQC, we performed a direct comparison with a fixed-order result focusing specifically on the problematic channel involving gluon splitting into a $q\bar{q}$ pair in the back-to-back limit ($z \to 1$).
We employed the parton-level generator \texttt{Event2}~\cite{Catani:1996vz} to obtain the NLO QCD prediction for this specific $g \to q\bar{q}$ contribution to the QQC as $z \to 1$. Crucially, the \texttt{Event2} calculation yields a finite result and independent of the IRC cutoff in the back-to-back limit, numerically confirming the anticipated IRC safety of the QQC in this configuration. 
Most importantly, the fixed-order result exactly matches the prediction obtained from our factorization theorem evaluated in the same limit. This agreement, illustrated in Figure~\ref{fig:event2_backtoback}, provides a highly non-trivial cross-check of our theoretical framework, demonstrating that our perturbative description based on the factorization theorem accurately captures the intricate cancellation of divergences and correctly predicts the perturbative result.

\begin{figure}
  \centering
  \includegraphics[width=0.49\textwidth,trim= 0 0 0 0.85cm, clip]{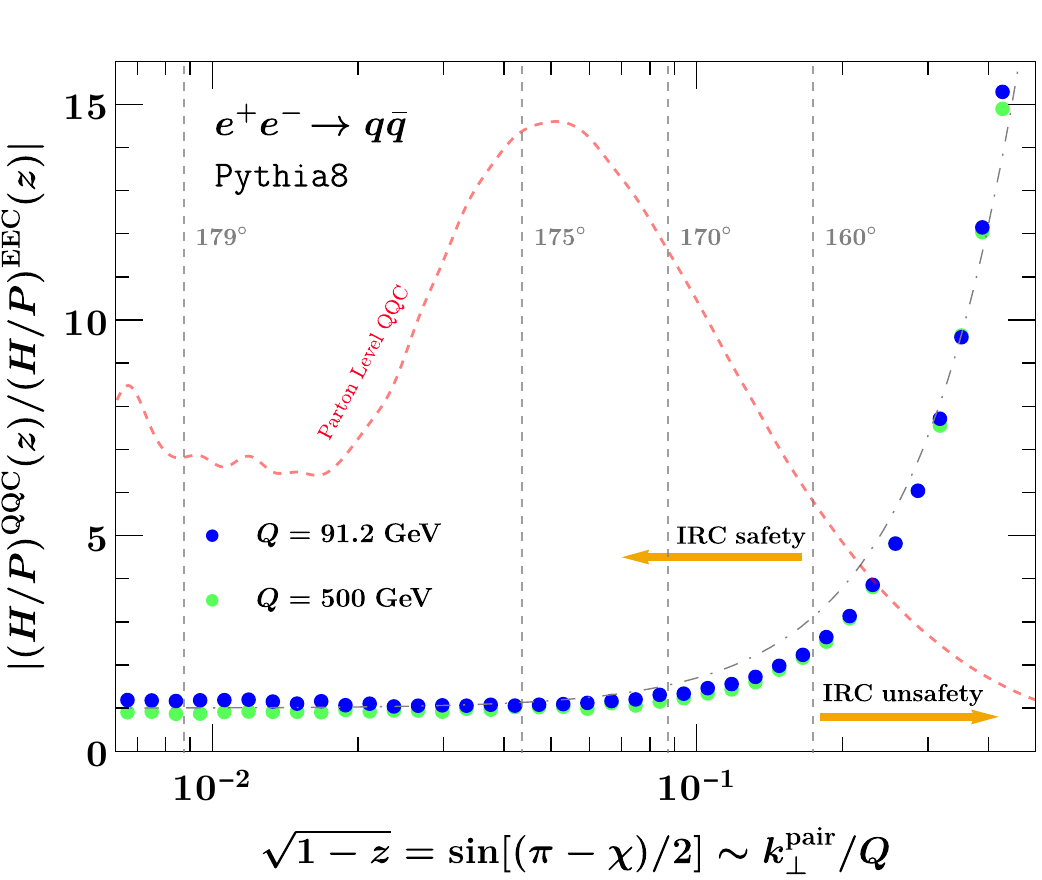} 
  \caption{Ratio of hadronization effects in QQC and EEC in the back-to-back limit. While for a generic angle the hadronization effects in the QQC are much larger than the ones for an IRC safe observable like the EEC, as we approach the back-to-back region we see a strong suppression of such effects in accordance with LP safety. The red dashed line depicts the parton level QQC to help visualize the IRC safe range of the observable. Finally, the gray dashed line corresponds to $y \sim (1-z) \sim (k_t/Q)^2$ indicating that the hadronization model indicates an hadronization corrections quadratic in the transverse momentum of the radiation. }
  \label{fig:hadronization}
\end{figure}

\paragraph*{Hadronization Effects---}
A further test of our results is the sensitivity of the back-to-back limit of QQC to hadronization effects. In \fig{hadronization}, we see that the hadronization effects for the QQC in the bulk are very large compared to the ones of an IRC safe observable like the EEC. However, they dramatically decrease as we approach the back-to-back limit, in line with the IRC safety of the QQC in this regime. In particular, we see that at center-of-mass energies corresponding to the Z peak, there is a ten-degree-wide angular region in which hadronization effects are comparable to the ones of the EEC. Although this might seem a narrow range, this window can be measured very precisely on charged tracks and thus has significant potential for phenomenological studies. As an example, a recent reanalysis of ALEPH data~\cite{Bossi:2025xsi} reports ${\cal O}(70)$ bins in this window for the tracks-based EEC. We leave a quantitative study of hadronization for future work.

\paragraph*{Conclusions.---}In this Letter, we have shown that the Charge-Charge Correlation, a fundamental probe of gauge theory dynamics known to be IRC unsafe, becomes perturbatively calculable in the back-to-back kinematic limit $z\to 1$. The IRC safety of the QQC in this asymptotic regime is a consequence of the suppression of the source of IRC divergences by powers of the angular variable $(1-z)$, a mechanism that we dub leading-power safety. At leading power, we have derived a SCET factorization theorem which does not rely on non-perturbative fragmentation functions. This factorization features a novel theoretical ingredient, the QQC jet function, which we determined to $\cO(\alpha_s^3)$. While nothing was known beyond $\cO(\alpha_s)$ in QCD for the QQC, our theoretical framework has allowed us to carry out the resummation for this observable to an unprecedented N$^4$LL accuracy. Our framework has been rigorously validated against analytic constraints from 
$\cN=4$ SYM and direct numerical calculations.
This work transforms the QQC from a incalculable, IRC unsafe observable into a high-precision phenomenological tool. It opens a new avenue for testing the intricate charge-flow dynamics of QCD and provides a powerful new observable for experimental analyses at $e^+e^-$ colliders. The generalization of our results to $ep$ and $pp$ collisions is straightforward following \cite{Gao:2019ojf,Li:2020bub,Li:2021txc}. Analogous considerations apply to correlators of other conserved global charges, that deserve a similar level of theoretical scrutiny in the future. By operating at the brink of IRC safety, we have unveiled a new, calculable window into the structure of quantum field theory.

\paragraph*{Acknowledgements.---}
We thank Hao Chen, Zhaoyan Pang, Emeri Sokatchev, and Alexander Zhiboedov for interesting discussions on the topic. 
PM and GV wish to thank the Center for High Energy Physics of Peking University for hospitality while this work was carried out. 
The work of PM is funded by the European Union (ERC, grant agreement No. 101044599).
Views and opinions expressed are however those of the authors only and do not necessarily reflect those of the European Union or the European Research Council Executive Agency. Neither the European Union
nor the granting authority can be held responsible for them.
The work of HXZ is funded by the National
Natural Science Foundation of China under contract No.
12425505 and The Fundamental Research Funds for the
Central Universities, Peking University.

\bibliographystyle{apsrev4-2}
\bibliography{refs}

\newpage

\onecolumngrid
\newpage
\appendix

\makeatletter
\renewcommand\@biblabel[1]{[#1S]}

\setcounter{equation}{0}
\makeatother

\section*{Supplemental Material}\label{app:supplemental}

\subsection{Renormalization Group Equations}

In this section, we summarize the renormalization group (RG) structure of the ingredients entering the factorization formula presented in \eqref{eq:QQC_fact_thm} of the main text. 
The solution to these RGEs for the hard and jet functions determines the resummed object entering \eq{QQC_resummed}. The logarithmic accuracy of the resummation is determined by the perturbative order at which various ingredients are computed. 
These include the hard and jet boundary terms, the cusp and non-cusp anomalous dimensions, the rapidity anomalous dimension, and the QCD beta function, which is currently known to 5 loops \cite{Herzog:2018kwj,Baikov:2016tgj}. 
Table~\ref{tbl:log_counting} summarizes the required orders of each quantity at different levels of logarithmic accuracy. 

The hard function $H_{q\bar{q}}$ and the jet function $\JQQC_q$ satisfy the following $\mu$-renormalization group equations (RGEs):
\begin{align} \label{eq:RGEs}
 \mu \frac{\df}{\df\mu}{\ln H_{q\bq}(Q,\mu)} &= \gamma^q_H(Q,\mu)\,, \nn\\
 \mu \frac{\df}{\df \mu} \ln \JQQC_q\left(b_T, \mu, \frac{Q b_T}{ \upsilon}\right) &=  \gamma_{\cJ_q}(\mu,\upsilon \mu/Q)\,.
\end{align}
Here, the anomalous dimensions decompose into cusp and non-cusp components as follows~\cite{Duhr:2022yyp}:
\begin{align} \label{eq:mu_anom_dims}
 \gamma^q_H(Q,\mu) &= 4 \, \GammaC^{q}[\as (\mu)] \ln\frac{Q}{\mu} + 4 \, \gamma^q_H[\as(\mu)]\,, \\
 \gamma_{\cJ_q}(\mu,\upsilon \mu/Q) &= 2 \, \GammaC^q[\as(\mu)] \ln\frac{\upsilon \mu}{Q} - 2 \, \gamma^q_H[\as(\mu)]\,.\nn
\end{align}
In Eq.~\eqref{eq:mu_anom_dims}, $\GammaC^q$ denotes the cusp anomalous dimension associated with lightlike Wilson lines in the fundamental representation~\cite{Korchemsky:1987wg,Bern:2005iz,Henn:2019swt}. From Table~\ref{tbl:log_counting} we see that its perturbative expansion to five loops is required for a N$^4$LL resummation, although this is currently unknown. However, its impact can be studied using the approximated value given in ref.~\cite{Herzog:2018kwj} and it is expected to be nearly negligible (cf.~ref.~\cite{Duhr:2022yyp}). The non-cusp term $\gamma^q_H[\as(\mu)]$ is extracted from the single poles of the quark form factor and is closely related to the quark collinear anomalous dimension~\cite{Agarwal:2021zft}. 
As is standard in SCET notation, we distinguish between the full anomalous dimensions and their non-cusp components by their argument: the former are functions of both a physical scale and a renormalization scale, while the latter depend only on $\as(\mu)$.

In addition to the $\mu$-RGEs, the jet function $\JQQC_q$ also satisfies a rapidity renormalization group equation:
\begin{align} \label{eq:RRGE}
 \upsilon \frac{\df}{\df \upsilon} \ln \JQQC_q\left(b_T, \mu, \frac{Q b_T}{ \upsilon}\right) = -\frac{1}{2} \gamma_r^q(b_T,\mu)\,.
\end{align}
The rapidity anomalous dimension $\gamma_r^q$ controls the evolution in rapidity and accounts for the logarithmic sensitivity to the large separation in rapidity of the two collinear sectors.
The rapidity anomalous dimension itself evolves with the following RGE,
\beq\label{eq:gammarRGE}
\mu \frac{\df}{\df \mu} \gamma_r^i(b_T,\mu) = -4  \GammaC^i[\as(\mu)]\,,
\eeq
whose solution reads 
\beq
	\gamma_r^i(b_T,\mu) = -4\int_{\mu_0}^\mu \frac{\df \mu^\prime}{\mu^\prime} \GammaC^i[\as(\mu')] +\gamma_r^i(\mu_0, b_T)\,,
\eeq
The rapidity anomalous dimension boundary $\gamma_r^i(\mu_0, b_T)$ has been calculated to 4 loops in Refs.~\cite{Moult:2022xzt,Duhr:2022yyp}.

\begin{table}
\centering
 \begin{tabular}{l|c|c|c|c|c} \hline\hline
  Accuracy & $H$, $\cJ$ & $\GammaC(\as)$  & $\gamma^q_H(\as)$ & $\gamma^q_r(\as)$ & $\beta(\as)$ \\\hline
  LL           & Tree level & 1-loop & --       & --       & 1-loop \\\hline
  NLL          & Tree level & 2-loop & 1-loop & 1-loop & 2-loop \\\hline
  NLL$^\prime$ & 1-loop     & 2-loop & 1-loop & 1-loop & 2-loop \\\hline
  NNLL         & 1-loop     & 3-loop & 2-loop & 2-loop & 3-loop \\\hline
  NNLL$^\prime$& 2-loop     & 3-loop & 2-loop & 2-loop & 3-loop \\\hline
  N$^3$LL         & 2-loop     & 4-loop & 3-loop & 3-loop & 4-loop \\\hline
  N$^3$LL$^\prime$& 3-loop     & 4-loop & 3-loop & 3-loop & 4-loop \\\hline
  N$^4$LL         & 3-loop     & 5-loop & 4-loop & 4-loop & 5-loop \\\hline
 \hline
 \end{tabular}
\caption{%
Logarithmic counting for the resummation accuracy in terms of the loop order of the ingredients: hard and jet functions ($H$, $\cJ$), the cusp anomalous dimension $\GammaC(\as)$, the quark anomalous dimension $\gamma^q_H(\as)$, the rapidity anomalous dimension $\gamma^q_r(\as)$, and the QCD beta function $\beta(\as)$.
}
\label{tbl:log_counting}
\end{table}

\end{document}